\documentclass{rQUF2e}

\usepackage{epstopdf}
\usepackage{subfigure}
\usepackage{lineno}
\usepackage{amsmath}
\usepackage{graphicx}

\newcommand{\corr}[2]{\mbox{Corr}\left[#1,#2\right]}
\usepackage[usenames, dvipsnames]{color}

\begin{document}


\title{On the interplay between multiscaling and stocks dependence}

\author{R. J. BUONOCORE$^{\dagger}$, G. BRANDI$^{\ast,\dagger}$,\thanks{$^\ast$Corresponding author. Email: giuseppe.brandi@kcl.ac.uk} R. N. MANTEGNA$^{\ddagger,\S,\P}$
and T. DI MATTEO$^{\dagger,\S,\P}$\\
\affil{$\dagger$ Department of Mathematics, King's College London, The Strand, London, WC2R 2LS, UK\\
       $\ddagger$ Dipartimento di Fisica e Chimica, Universit\`a degli Studi di Palermo, Viale delle Scienze, Ed. 18, I-90128 Palermo, Italy\\
       $\S$ Department of Computer Science, University College London, Gower Street, London, WC1E 6BT, UK\\
       $\P$ Complexity Science Hub Vienna, Josefstaedter Strasse 39, A 1080 Vienna, Austria}}

\maketitle

\begin{abstract}
We find a nonlinear dependence between an indicator of the degree of multiscaling of log-price time series of a stock and the average correlation of the stock with respect to the other stocks traded in the same market. This result is a robust stylized fact holding for different financial markets. We investigate this result conditional on the stocks' capitalization and on the kurtosis of stocks' log-returns in order to search for possible confounding effects. We show that a linear dependence with the logarithm of the capitalization and the logarithm of kurtosis does not explain the observed stylized fact, which we interpret as being originated from a deeper relationship.
\end{abstract}


\begin{classcode}G19, G10
\end{classcode}

\section{Introduction}
Financial time-series are characterized by the presence of so called stylized facts \citep{ramacont_review,chakraborti_review}. The most famous ones are: power law tails \citep{mantegna_stanley}, volatility clustering \citep{engle}, multiscaling \citep{mantegna_stanley,mantegna_stanley_book,lux_marchesi,dacorogna_book,calvet1,lux1,tiziana_dacorogna2,scaling_review_tiziana,reviewzhou} and the presence of a dependency structure between stocks \citep{mantegna1999,borghesi2007,aste2010,tumminello2010,musmeci2015,musmeci2016}. Stylized facts deserve attention due to the hidden data information they provide to risk and asset managers.
Power law tails, volatility clustering and multiscaling refer to univariate properties of financial time series while the dependency structure between stocks is a multivariate property.

The multiscaling property of financial time series has been widely studied in the last two decades. The research on this topic developed either on the theoretical side \citep{mandelbrot1,bacry1,calvet1,bacry_other_models,bacry_skewed} or on the empirical one \citep{mantegna_stanley,mantegna_stanley_book,dacorogna_book,calvet1,scaling_review_tiziana,bartolozzi1,bartolozzi2,zhou2,liu1,morales1,zhou3,kristoufek}. 
On the other hand, the dependency structure of the markets has been observed across different industries and asset classes \citep{mantegna_stanley_book, musmeci2015_2}. The study of stocks dependence became widespread since the introduction of Markovitz's modern portfolio theory \citep{markowitz}, in which the optimization method focuses on the role of the covariance (or correlation) matrix of stock returns.\\ From an empirical point of view, one may be interested on the portfolio construction basing the optimization algorithm not only on the observed dependency structure but also the degree of multiscaling each stock shows. In this case, one can either try to retain stocks with low level of multiscaling relying on the theoretical aspects of Brownian motions or to exploit the multiscaling feature of some stocks. This would lead to a new way of building portfolios, putting together multiscaling and dependency between stocks. 
However, despite the paramount importance of those two key stylized facts, no research has been produced to investigate the relationship between them.  
In this paper we find a new stylized fact showing a robust statistical relationship between the multiscaling property of log-prices time series of a stock and the average correlation of the stock with respect to the other stocks traded in the same financial market. We verify that this relationship holds in several leading stock markets and we investigate about its origin, providing empirical evidence that the result is not explained by the capitalization of the stocks or by its kurtosis coefficient, but rather from a deeper, intrinsic relationship. Theoretical attempts in similar directions can be found in \cite{mmrw} and \cite{morales3}. However, empirical evidence has been lacking so far. It is worth noting that \cite{micciche1} goes in the same spirit of this paper. In fact, the author observes a relationship between the volatility clustering in time and the correlation of volatility of each stock with volatility of other stocks traded in the same market.

The paper is organized as follows: in Sec. \ref{sec_mul_corr_proxy} we describe the tools and the dataset we use to perform the analysis, in Sec.\ref{sec_relationship} we present main results and in Sec.\ref{sec_conclusion} we draw our conclusions.

\section{Methods and dataset of analysis}\label{sec_mul_corr_proxy}
In this section we describe the methods used to estimate the univariate and multivariate properties of the time series and the dataset we use to perform our empirical analyses. Let us first fix the notation by defining the prices time series as $p(t)$ and the log-returns\footnote{In the text we refer to log as the natural logarithm.} over a $\tau$ time aggregation as \[r(t)=\ln\left[p(t+\tau)/p(t)\right]\] where $\tau$ is expressed in days in this paper. In what follows we work with log-returns time-series of zero mean obtained by removing the sample mean.
\subsection{Multiscaling and correlation proxies}
The multiscaling property of a financial time series \citep{scaling_review_tiziana} is detected by quantifying the non-linearity of the scaling exponents of the $q$-order moments of the log-returns' absolute value. In particular, the process $\ln[p(t)]$ is said to be multiscaling if
\begin{equation}\label{mult_def}
E\left[|r(t)|^q\right]=K(q)\tau^{qH(q)},
\end{equation}
where $K(q)$ is the $q$-moment for $\tau=1$, $H(q)$ is the so called Generalised Hurst Exponent which is a function of $q$ and the function $qH(q)$ is concave \citep{mandelbrot1} and codifies the scaling exponents of the process. A process is uniscaling when the function $H(q)$ does not depende on $q$, \textit{i.e.} $H(q)=H$ \citep{scaling_review_tiziana}, while multiscaling otherwise.  In light of Eq.\ref{mult_def} a possible way to define a multiscaling proxy is by quantifying the degree of non-linearity of the function $qH(q)$. In order to do so, first the scaling exponents $qH(q)$ have to be computed, which is done via a linear regression in log-log scale of Eq.\ref{mult_def}, which reads as
	\begin{equation}\label{log_mult_def}
	\ln\left(E\left[|r(t)|^q\right]\right)=qH(q)\ln(\tau) + \ln\left(K(q)\right),
	\end{equation}
	where $\tau$ in our analysis is taken in the range $\tau=[\tau_{min}, \tau_{max}]=[1,19]$ \citep{scaling_review_tiziana}. A multiscaling proxy can be obtained by fitting the measured scaling exponent with a second degree polynomial \citep{buonocore} of the form
\begin{equation}\label{mult_proxy}
\zeta(q)=qH(q)=Bq^2+Cq+D.
\end{equation}
To be consistent with the prescription of \citep{mandelbrot1,buonocore2}, Eq.\ref{mult_proxy} must satisfy some conditions, in particular:

\begin{equation}\label{mult_proxy_conditions}
\begin{aligned}
 \zeta(0)=0 \\
 \zeta(2)=1 \\
 \zeta(q)''<0.
\end{aligned}
\end{equation}

The first condition is a consequence of the definition of scaling exponent. The second one comes from the fact that returns are expected to be uncorrelated at daily frequency \citep{ramacont_review,chakraborti_review}. The third condition is a concavity requirement. Solving for these conditions, Eq.\ref{mult_proxy} becomes:

\begin{equation}\label{mult_proxy2}
qH(q)=Bq^2+\bigg(\frac{1}{2}-2B\bigg)q=Bq^2+Aq.
\end{equation}

In this statistical setting, $\hat B$ represents the non-linearity proxy. If $\hat B=0$, the process is uniscaling, while if $\hat B\neq0$, the process is multiscaling (cfr. \citep{buonocore}). Together with $\hat B$, also $\hat A$ gives information about the analysed process. In particular, when $\hat B<0$, we expect, by Eq.\ref{mult_proxy2}, that $\hat A>0.5$ for concavity. When $\hat B\approx0$, we expect $\hat A\approx0.5$.\\
As explained in the introduction, we are going to associate the multiscaling property (through the multiscaling proxy $B$) with a multivariate quantity which provides information regarding the interrelation between stocks movements.
Co-movements between stocks is a fundamental concept in Finance. It helps understanding the behaviour of stocks and building efficient portfolios. Furthermore, in specific financial conditions, like distress periods, association between stocks reveals more information about risk variations than other standard measures, eg. variance. Various forms of association between stocks are available from the statistical toolbox. In financial economics, the reference measure for association is the Pearson's correlation.
Given its easiness of interpretation and popularity to quantify the dependency structure between stocks, we consider the average of the Pearson's correlation coefficient $\bar\rho_{i}$ between the $i$th stock and all the other stocks traded in the same financial market. Taking the Pearson's correlation coefficient between stock $i$ and $j$ as:  
\begin{equation}\label{pearson}
\displaystyle\rho_{ij}=\corr{r_i}{r_j}=\frac{E[r_ir_j]-E[r_i]E[r_j]}{\sqrt{Var[r_i]}\sqrt{Var[r_j]}}, 
\end{equation}
the average Pearson's correlation coefficient, $\bar\rho_{i}$ is defined as:
\begin{equation}\label{stdproxy}
\bar{\rho}_i=\frac{1}{N-1}\sum_{j\neq i=1}^N\rho_{ij},
\end{equation}
where $N$ is the number of stocks in a given financial market. This measures a proxy on how a stock relates to all the other stocks in the market and, in combination with the scaling property, it can give valuable information to the risk and asset managers.
\subsection{Dataset}
The data we use for our analyses is made of six different sets of stocks. In particular we investigate time-series of the \textit{London Stock Exchange} (LSE), \textit{Frankfurt Stock Exchange} (FWB), \textit{Tokyo Stock Exchange} (TSE) and \textit{Hong Kong Stock Exchange} (HKSE). To these four sets of data we add another set of data obtained by merging stocks traded at the \textit{New York Stock Exchange}, at the \textit{Nasdaq Stock Market}, and at the \textit{NYSE MKT LLC}. We name this dataset as NYSE17. For these five sets of data, we consider the closure price of stocks recorded on a daily basis from the 03/01/2000 up to the 12/05/2017. A last dataset comprises the closure price recorded on a daily basis of the stocks traded at the \textit{New York Stock Exchange} from the 02/01/1985 to the 31/12/1999. We identify this dataset with the acronym NYSE99. Within the former (latter) time period we chose to consider only the stocks traded throughout the whole period of time, with an Initial Public Offer date earlier than the 03/01/2000 (02/01/1985) and traded at least up to the 15/07/2016 (31/12/1999). 
%

Moreover, since our aim is to perform a correlation analysis, our dataset cannot be used as it is. Indeed, price time series are not always present for all stocks due to the fact the some stocks have not been traded on certain days. In order to overcome this issue we apply a data cleaning procedure which allows to retain as many stocks as possible. The pre-processing procedure is based on dragging the last available value when the following one is missing. Details of this procedure can be found in Appendix \ref{appendix_cleaning}, while a summary of the number of stocks which are included in the dataset of each market after the pre-processing is presented in Tab.\ref{dataset_after}.
\begin{table}[!h]
\begin{center}
\begin{tabular}{|c|c|}
\hline
Market	& \# of stocks\\
\hline
NYSE17			&	$1202$\\
\hline
LSE				&	$144$\\
\hline
FWB				&	$126$\\
\hline
TSE				&	$724$\\
\hline
HKSE			&	$340$\\
\hline
NYSE99			&	$313$\\
\hline
\end{tabular}
\caption{Summary of the number of stocks of each set of data obtained after the application of the pre-processing procedure described in Appendix \ref{appendix_cleaning}.}
\label{dataset_after}
\end{center}
\end{table}
As for the capitalization, we consider the capitalization time series for each stock and we take the median capitalization of each stock over the considered time interval. We chose the median and not the mean because we want to keep the most representative value of the capitalization over the given time period. For few stocks the capitalization was not available in the source database. In this case, when a capitalization analysis is involved we simply remove stocks without information about it.

\section{Multiscaling and average correlation}\label{sec_relationship}
\subsection{Empirical evidence}\label{subsec_main}
\begin{figure}
\begin{center}
\includegraphics[width=0.70\textwidth]{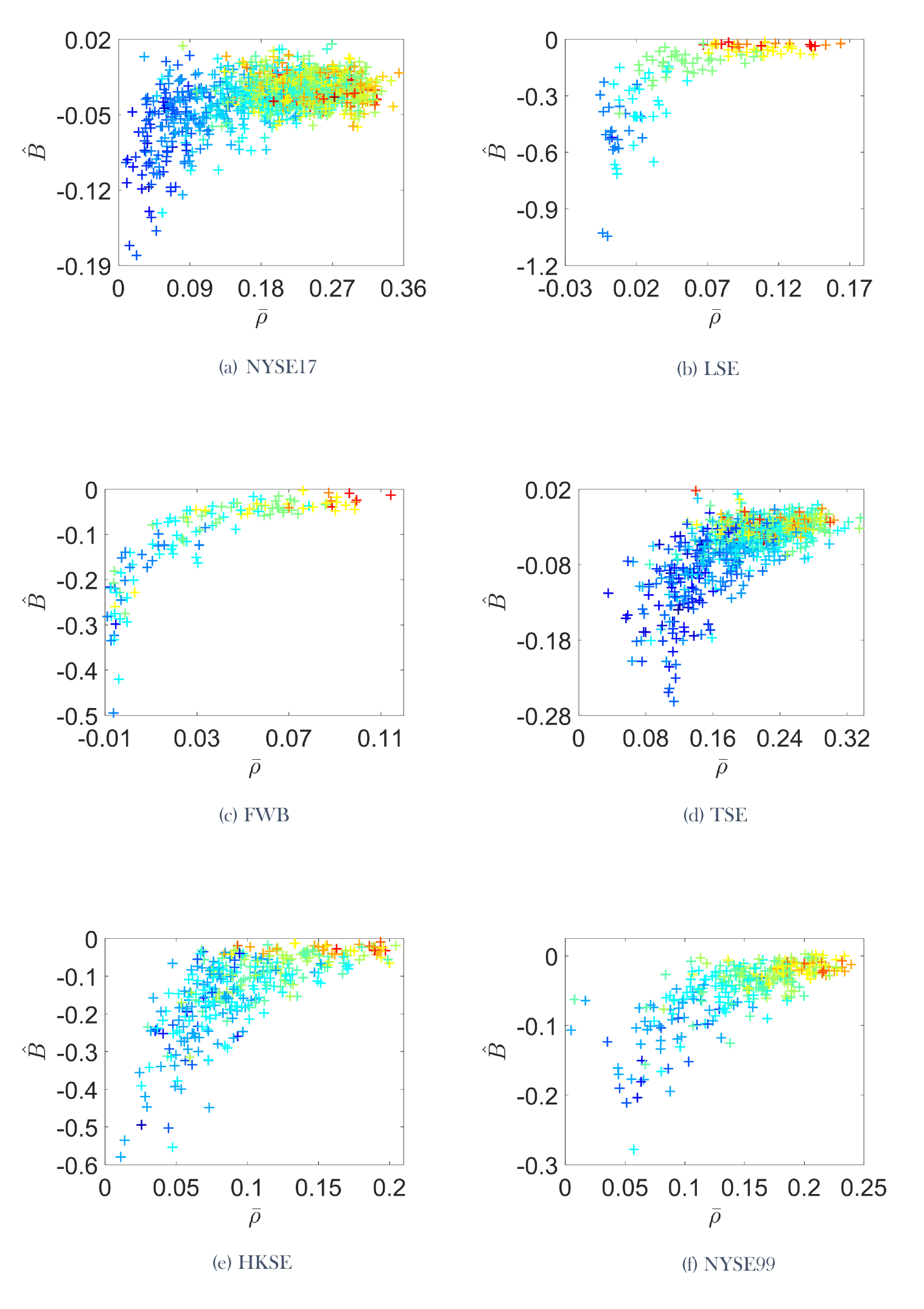}
\end{center}
\caption{Empirical evidence of the dependence between the degree of multiscaling measured by the proxy $\hat B$ of the log-price financial time series and its average correlation $\bar \rho$ with the other stocks traded in the same market. The colours from blue to red represent an increasing capitalization.}
\label{B_VS_corr_plain_market_cap}
\end{figure}

In this section we present the empirical evidence of a new stylized fact. We find an empirical relationship between the scaling property of a financial time series and the average correlation of stock with other stocks traded in the same market. Specifically, Fig.\ref{B_VS_corr_plain_market_cap} shows the scatter plot of $\hat B$ with respect to $\bar \rho$. The range of the parameter $\tau$ over which the scaling is computed is chosen to be $\tau=[\tau_{min},\tau_{max}]=[1,19]$ \citep{tiziana_dacorogna,tiziana_dacorogna2}, since an approximate linearity is present in this region. The range of $q$ is set as $q\in[0.1,1]$, with steps of $0.1$, following the prescription used in \citep{buonocore}. The analysis has been performed for different values of $\tau_{max}$ and $q$ for which the results remains stable. In Fig.\ref{B_VS_corr_plain_market_cap}, the colour of the dots indicates the capitalization with increasing value ranging from dark blue to red in a log scale. since the interval in capitalization between the most capitalized stocks and the lowest capitalized one is covering many orders of magnitudes. As it can be seen from the scatter plots of Fig.\ref{B_VS_corr_plain_market_cap}, a nonlinear relationship between $\hat B$ and $\bar \rho$ emerges across all markets.\footnote{We carried out the analysis also using MFDFA \citep{reviewzhou}, and the results are qualitatively equivalent, suggesting the analysis is not method dependent.} To display the capitalization we use a log value. It is also worth pointing out that there is a monotonic relationship between the market capitalization and both the average correlation and the degree of multiscaling $\hat{B}$. In order to make a quantitative assessment about these (nonlinear) relationship, we use the Kendall $\tau$ ($K\tau$) rank correlation \citep{hollander} between the two variables. For a generic set of variables $x$ and $y$, $K\tau$ is defined as

\begin{equation}\label{kendall}
K\tau_{x,y}=\frac{2}{n(n-1)}\sum_{k < w}sgn(x_k-x_w)sgn(y_k-y_w),
\end{equation}

where $k$ and $w$ are the rank positions of the variables and $sgn(z)$ is the sign function which equals $1$ if the sign of $z$ is positive and $-1$ if the sign of $z$ is negative. This measure is particularly useful for nonlinear relationships since it computes the similarity based on the ranks of the data.
Results of the analysis are reported in Tab.\ref{B_VS_rho} and reveal a positive and significant relationship between the two stylized facts, confirming quantitatively the pattern shown in Fig.\ref{B_VS_corr_plain_market_cap}. The next section is dedicated to the investigation of the robustness of this result.   

\begin{table}
\begin{center}
\begin{tabular}{|c|c|}
\hline 
Market & $K\tau$  \\ 
\hline 
NYSE17 & $0.30$  \\ 
\hline 
LSE & $0.65$ \\ 
\hline 
FWB & $0.74$  \\ 
\hline 
TSE & $0.46$  \\ 
\hline 
HKSE & $0.54$  \\ 
\hline 
NYSE99 & $0.53$  \\ 
\hline 
\end{tabular}
\caption{Kendall $\tau$ ($K\tau$) correlation between the multiscaling proxy $\hat B$ and $\bar \rho$. All the p-values are less than 0.01.}
\label{B_VS_rho}
\end{center}
\end{table}

\subsection{Robustness of empirical results}
\cite{buonocore} found that the scaling of the moments at a low aggregation horizon is strongly affected by bias with respect to the expected asymptotic behaviour. The bias comes either from the presence of the power law tails of log-returns probability density function or from the presence of the memory of the log-return time series. Hereafter, we want to understand whether the behaviour displayed in Fig.\ref{B_VS_corr_plain_market_cap} is primarily due to the power law tails or to the time memory of log-returns. In order to uncover the origin of the multiscaling property reported in Fig.\ref{B_VS_corr_plain_market_cap}, we employ two different procedures to generate surrogate data, namely: 

\begin{enumerate}
\item[1.] Shuffle the $r(t)$ time-series in every market in a synchronous way across different stocks;
\item[2.] Use the normalization methodology of \cite{zhou} to change the shape of the distribution.
\end{enumerate}

\begin{figure}
\begin{center}
\includegraphics[width=0.70\textwidth]{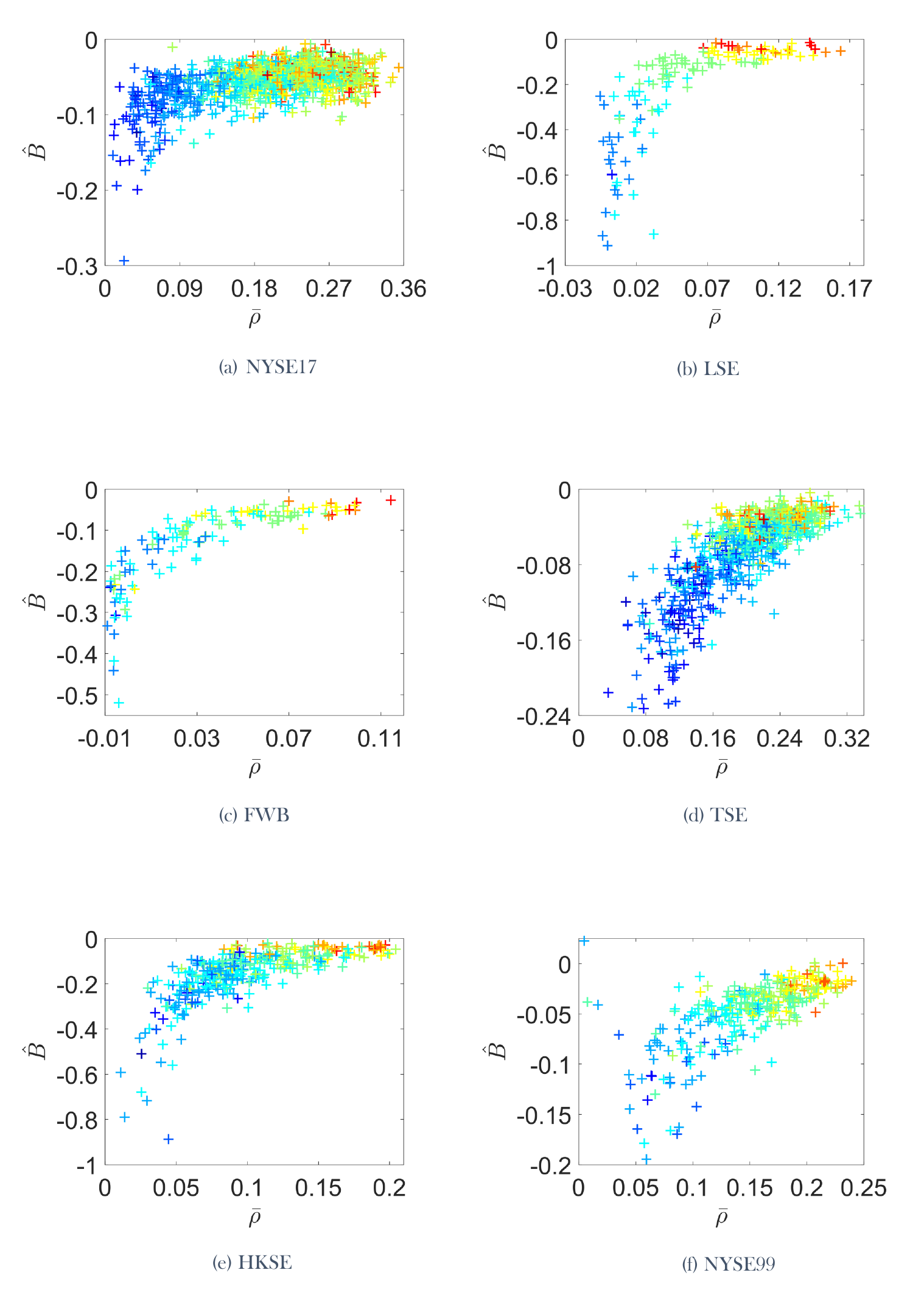}
\end{center}
\caption{Empirical evidence of the dependence between the degree of multiscaling measured by the proxy $\hat B$ of the log-prices financial time series when the log-returns are shuffled and its average correlation $\bar \rho$ with the other stocks traded in the same market. The colours from blue to red represent the increasing capitalization.}
\label{fig_B_VS_corr_plain_market_cap_shuffled}
\end{figure}

The first transformation is used so that the autocorrelation structure of each time series is destroyed whereas the correlation structure is preserved. In Fig.\ref{fig_B_VS_corr_plain_market_cap_shuffled} we show the relationship between $\hat B$ and $\hat \rho$, computed on shuffled time series. In Tab.\ref{tab_B_VS_rho_shuffled} we report the Kendall $\tau$ correlation up to two significant digits between the two quantities. It appears evident, from both a graphical and quantitative analysis, that the relationship still holds almost unchanged after shuffling.
\begin{table}
\begin{center}
\begin{tabular}{|c|c|}
\hline 
Market & $K\tau$  \\ 
\hline 
NYSE17 & $0.30$ \\ 
\hline 
LSE & $0.66$  \\ 
\hline 
FWB & $0.74$  \\ 
\hline 
TSE & $0.58$ \\ 
\hline 
HKSE & $0.59$ \\ 
\hline 
NYSE99 & $0.52$  \\ 
\hline 
\end{tabular} 
\end{center}
\caption{Kendall $\tau$ ($K\tau$) correlation between $\hat B$ and $\bar \rho$ when the log-returns are shuffled but preserving the correlation. All the p-values are less than 0.01.}
\label{tab_B_VS_rho_shuffled}
\end{table}

In order to preserve the memory structure of the time series but removing the effect of the shape of log-returns distribution, we use the normalization technique discussed in \citep{zhou,buonocore}. It consists of
\begin{enumerate}
		 
		\item[1.] Ranking the empirical returns according to their values;
		\item[2.] Recording the sequence of ranks;
		\item[3.] Extracting random numbers from the desired unconditional distribution;
		\item[4.] Ordering them following the sequence or ranks of the empirical one.
\end{enumerate}
We employ this procedure by using as desired unconditional distribution the Normal one. We limit ourself to report here that, once the shape effects are removed, the dependence is completely destroyed. 

\begin{figure}
\begin{center}
\includegraphics[width=0.70\textwidth]{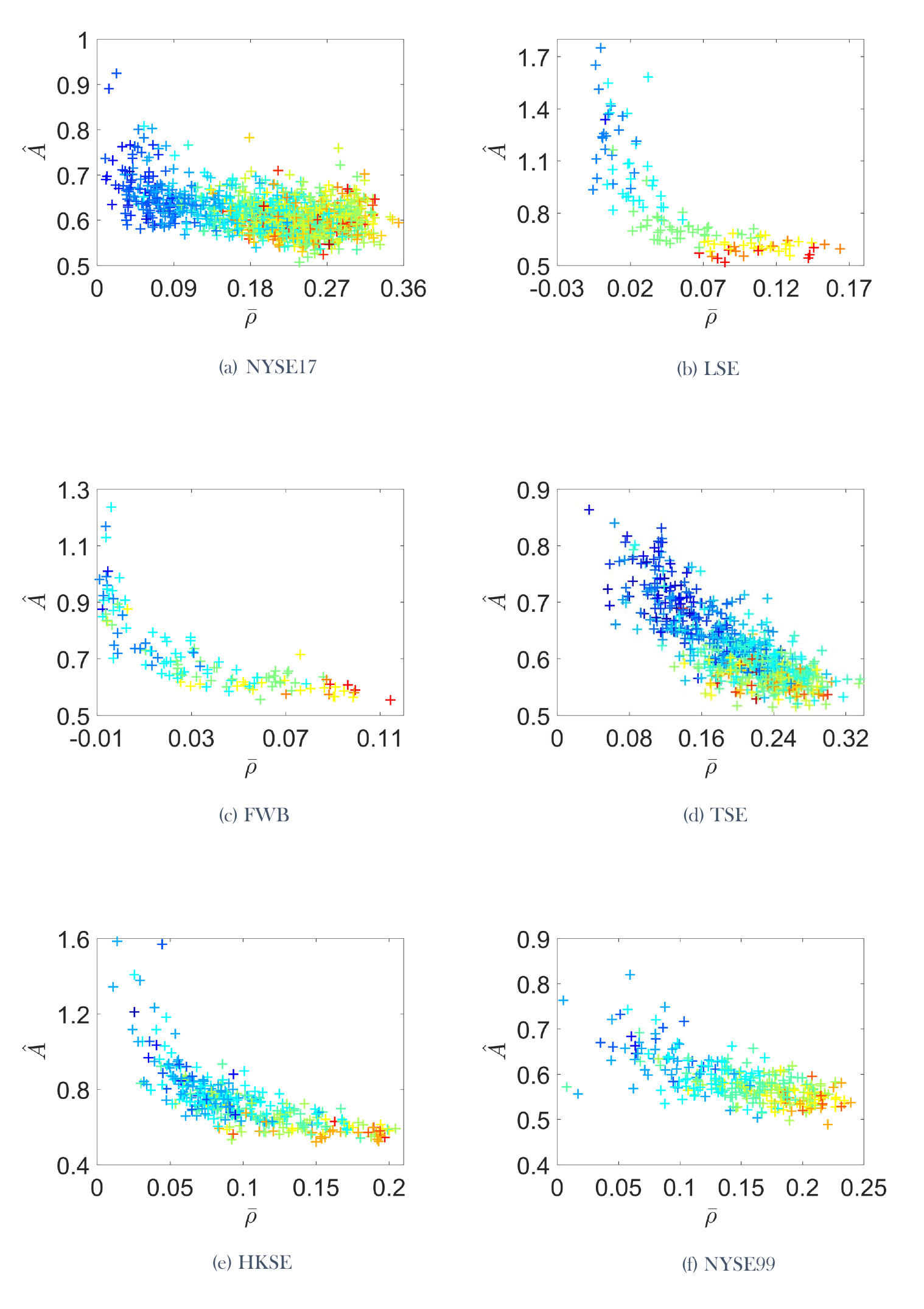}
\end{center}
\caption{Empirical evidence of the dependence between the multiscaling properties, represented by the proxy $\hat A$ of a time-series and its average correlation $\bar \rho$. The colours from blue to red represent the increasing capitalization.}
\label{A_VS_corr_plain_market_cap}
\end{figure}

Since those results suggest that fat tails are the main contribution to the multiscaling indicator we empirically find, we should observe that the value of $\hat A$ (cfr. Eq.\ref{mult_proxy}) converges to $0.5$ for stocks with $\hat B\approx0$ for consistency. This observation is confirmed by plotting the scatter plot of $\hat A$ against $\bar \rho$ as we report in Fig.\ref{A_VS_corr_plain_market_cap}. In fact, this figure shows that $\hat A$ tends to $0.5$ when $\hat B$ assumes values near zero, i.e. for highly capitalized stocks.
\begin{table}
\begin{center}
\begin{tabular}{|c|c|}
\hline 
Market & $K\tau$ \\ 
\hline 
NYSE17 & $-0.31$  \\ 
\hline 
LSE & $-0.64$  \\ 
\hline 
FWB & $-0.34$  \\ 
\hline 
TSE & $-0.43$  \\ 
\hline 
HKSE & $-0.54$ \\ 
\hline 
NYSE99 & $-0.49$  \\ 
\hline 
\end{tabular}
\caption{Kendall $\tau$ ($K\tau$) correlation between the multiscaling proxy $\hat A$ and $\bar \rho$. All the p-values are less than 0.01.}
\label{A_VS_rho}
\end{center}
\end{table}

\subsection{Role of the capitalization}\label{subsec_cap}
\begin{figure}[!h]
\begin{center}
\includegraphics[width=0.70\textwidth]{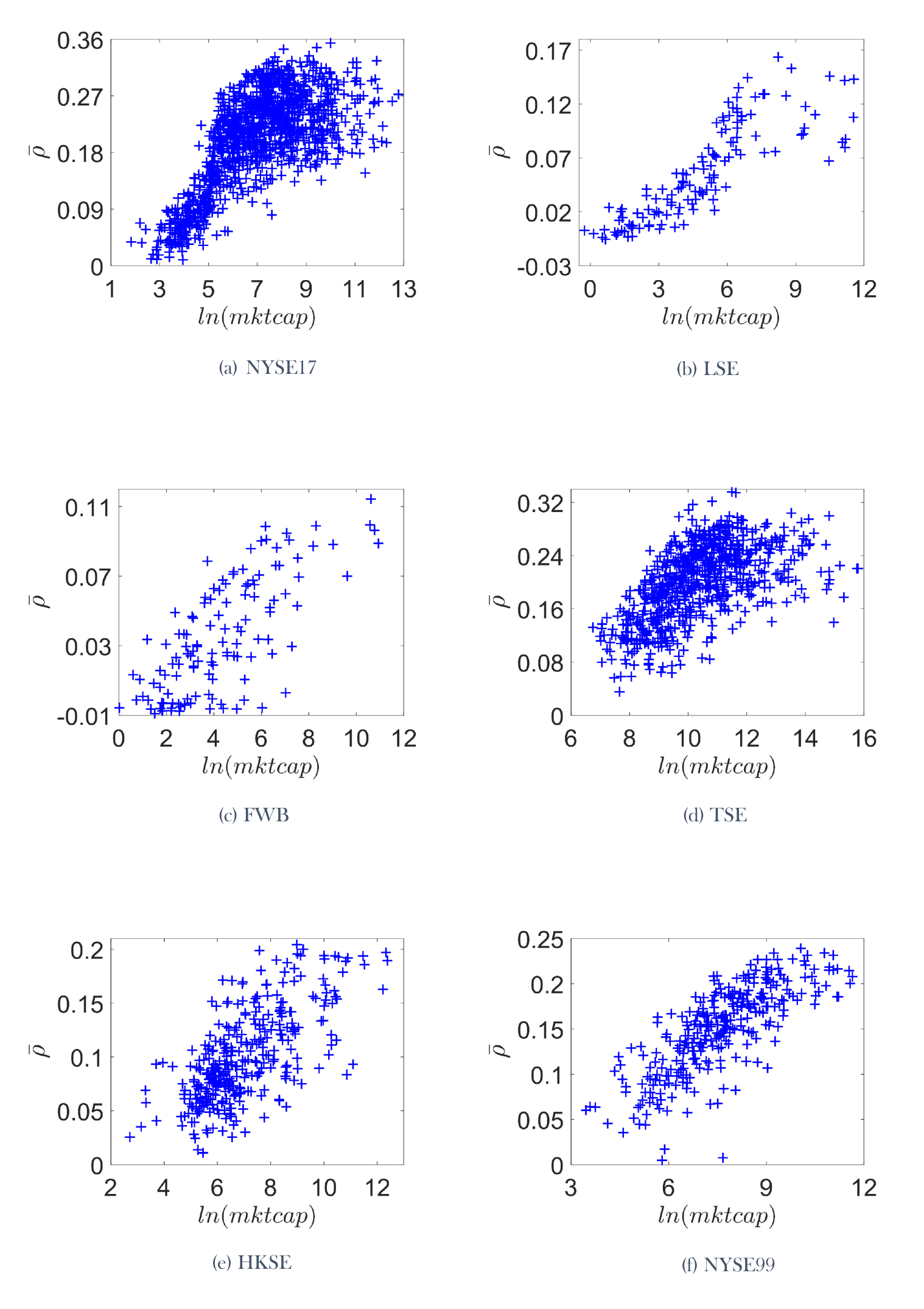}
\end{center}
\caption{ Empirical evidence of the dependence between the logarithm of the capitalization $\ln(mktcap)$ of stocks and the average correlation $\bar \rho$ for the six markets.}
\label{rho_VS_market_cap}
\end{figure}

\begin{figure}[!h]
\begin{center}
\includegraphics[width=0.70\textwidth]{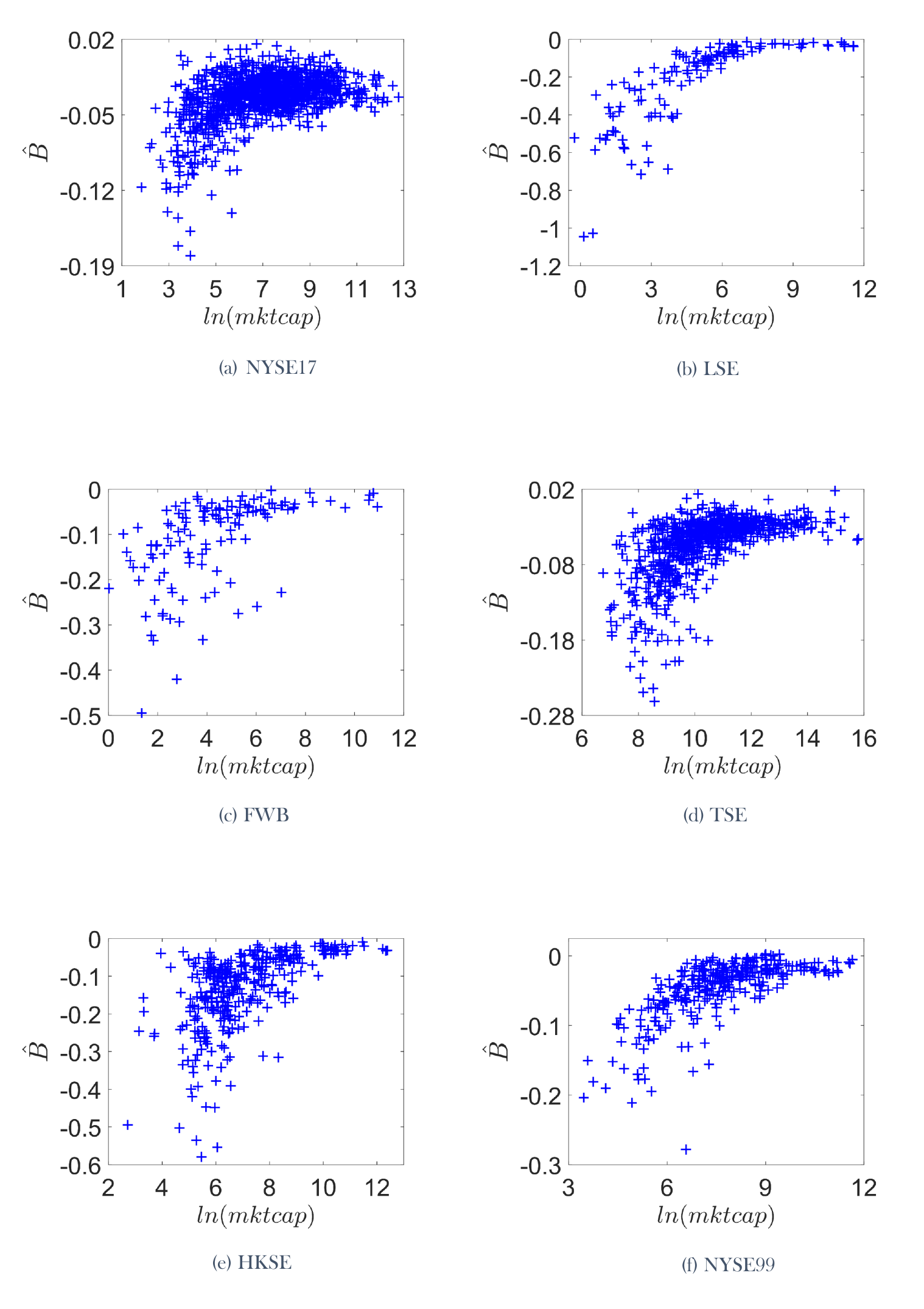}
\end{center}
\caption{ Empirical evidence of the dependence between the logarithm of the capitalization $\ln(mktcap)$ of stocks and the multiscaling proxy $\hat B$ for the six markets.}
\label{B_VS_market_cap}
\end{figure}

As pointed out in Subsec.\ref{subsec_main}, the scatter plots of Fig.\ref{B_VS_corr_plain_market_cap} suggest that $\hat B$ and $\bar \rho$ depend on the logarithm of capitalization, $\ln(mktcap)$. To confirm this observation and to quantify the effect, we plot in Figs.{\ref{rho_VS_market_cap} and \ref{B_VS_market_cap}} the logarithm of capitalization with respect to 
$\bar \rho$ and to $\hat B$ respectively, for the six datasets. We report in Tab.\ref{tab_B_ln_cap} the Pearson's correlation $\rho$ between $\hat B$ and the logarithm of capitalization and in Tab.\ref{tab_rho_bar_ln_cap} the Pearson's correlation between $\bar \rho$ and the logarithm of capitalization.
\begin{table}[!h]
\begin{minipage}{0.40\textwidth}
\begin{center}
\begin{tabular}{|c|c|}
\hline 
Market & $\rho$ \\ 
\hline 
NYSE17 & $0.29$  \\ 
\hline 
LSE & $0.73$ \\ 
\hline 
FWB & $0.48$  \\ 
\hline 
TSE & $0.45$  \\ 
\hline 
HKSE & $0.47$ \\ 
\hline 
NYSE99 & $0.53$ \\ 
\hline 
\end{tabular}
\end{center}
\caption{Pearson's correlation coefficient between the multiscaling proxy $\hat B$ and the logarithm of capitalization. All the p-values are less than 0.01.}
\label{tab_B_ln_cap}
\end{minipage}
\qquad
\begin{minipage}{0.40\textwidth}
\begin{center}
\begin{tabular}{|c|c|}
\hline 
Market & $\rho$  \\ 
\hline 
NYSE17 & $0.69$ \\ 
\hline 
LSE & $0.83$ \\ 
\hline 
FWB & $0.74$  \\ 
\hline 
TSE & $0.58$  \\ 
\hline 
HKSE & $0.67$  \\ 
\hline 
NYSE99 & $0.77$  \\ 
\hline 
\end{tabular} 
\end{center}
\caption{Pearson's correlation coefficient between the average correlation $\bar \rho$ and the logarithm of capitalization. All the p-values are less than 0.01.}
\label{tab_rho_bar_ln_cap}
\end{minipage}
\end{table}
Tabs.\ref{tab_B_ln_cap} and \ref{tab_rho_bar_ln_cap} show that the correlation is quite pronounced in both cases. Given this result, we want to verify if the dependence between the average correlation and the multiscaling proxy is driven only by the logarithm of capitalization. In order to assess this we compute the partial correlation \citep{parcorr} between $\bar \rho$ and $\hat B$ using as control variable the logarithm of capitalization. Partial correlation between two generic variables $X$ and $Y$ controlling for variables in $\textbf{Z}$ writes $\rho_{X,Y|Z}$ and can be calculated following a two steps approach. In the first step, we run a linear regression between each variable of interest against $\textbf{Z}$, which reads 	
	\begin{equation}\label{step_1_parcorr}
	\begin{aligned}
	X=\beta_0+\beta_1\textbf{Z}+\epsilon_X\\
	Y=\gamma_0+\gamma_1\textbf{Z}+\epsilon_Y,
	\end{aligned}
	\end{equation}
	
where the  $\beta$s and $\gamma$s are parameters and $\epsilon_X$ and $\epsilon_Y$ are the regressions residual, which account for the information not explained by $\textbf{Z}$. In the second step of the computation, the Pearson's correlation of Eq.\ref{pearson} is calculated between $\epsilon_X$ and $\epsilon_Y$, i.e.:

 		\begin{equation}\label{step_2_parcorr}
 		 \rho_{X,Y|\textbf{Z}}=\rho_{\epsilon_X\epsilon_Y}=\frac{E[\epsilon_X\epsilon_Y]-E[\epsilon_X]E[\epsilon_Y]}{\sqrt{Var[\epsilon_X]}\sqrt{Var[\epsilon_Y]}},
 			\end{equation}
	

where $\rho_{X,Y|\textbf{Z}}$ is the correlation between $X$ and $Y$ removing the effect of $\textbf{Z}$ from both variables. In our analysis, $X=\hat B$, $Y=\bar \rho$ and  $\textbf{Z}=\ln(mktcap)$. 

We report in Tab.\ref{tab_par_corr} the results of this analysis along with the $R^2$, which is the coefficient of determination \citep{r2} of the linear fit between $\bar \rho$ and $\hat B$ and the logarithm of the capitalization, defined as the amount of variation of the dependent variable explained by the independent variables, which is an estimate of the goodness of the fit. As we can see the correlation remains significant even after the removal of the control variable, meaning that the interplay between scaling and correlation has a deeper origin than just a linear relationship between the two variables and the logarithm of capitalization.
\begin{table}[!h]
\begin{center}
\begin{tabular}{|c|c|c|c|}
\hline 
Market & $\rho_{par}$ & $R^2_{\bar \rho}$ & $R^2_{\hat B}$ \\ 
\hline 
NYSE17 & $0.34$  & $0.47$ & $0.22$ \\ 
\hline 
LSE & $0.25$ &  $0.69$ & $0.57$ \\ 
\hline 
FWB & $0.68$ &  $0.54$ & $0.30$ \\ 
\hline 
TSE & $0.52$ & $0.33$ & $0.33$ \\ 
\hline 
HKSE & $0.50$ &  $0.45$ & $0.32$ \\ 
\hline 
NYSE99 & $0.44$ & $0.60$ & $0.46$ \\ 
\hline 
\end{tabular} 
\end{center}
\caption{Partial Pearson's correlation $\rho_{par}= \rho_{\hat B,  \bar \rho |\ln(mktcap)}$ between the average correlation $\bar\rho$ and $\hat B$. All the p-values are less than 0.01. The coefficients of determination $R^2_{\bar \rho}$ and $R^2_{\hat B}$ are also reported for the linear fit between the logarithm of the capitalization and $\bar\rho$ and $\hat B$ respectively.}
\label{tab_par_corr}
\end{table}

In order to be able to directly compare the results depicted in Tab.\ref{B_VS_rho} regarding relationship between $\bar\rho$ and $\hat B$ measured by $K\tau$ with the level of association of the two variables conditional to the capitalization, we also computed the $K\tau$ correlation between the two variables after the linear relationship between the two variables and the capitalization is filtered out. We name this linear-partial $K\tau$ correlation in order to discriminate from the partial $K\tau$ correlation which implies a different type of partialling out. This is equivalent to compute the Kendall correlation $K\tau$ between the residuals found in Eq.\ref{step_1_parcorr}.

 The results are reported in Tab.\ref{tab_par_tau} and show that the relationship, even if attenuated, is still present. In order to better control for non-linear relationships between capitalization and the two variables, we also conditioned the correlation between $\bar\rho$ and $\hat B$ on the squared values of the logarithm of capitalization. The results of this further analysis show that the statistical significance of the partial correlation remains virtually unchanged.\footnote{Filtering out both linear and non-linear dependence via second degree polynomial regression, the dependence structure remains statistically significant in all markets apart the London Stock Exchange.}  
This result suggests that the relationship between multiscaling and average stocks correlation has a deeper origin.  

\begin{table}
\begin{center}
\begin{tabular}{|c|c|c|c|}
			\hline 
			Market & $K\tau_{par}$ & $R^2_{\bar \rho}$ & $R^2_{\hat B}$ \\ 
			\hline 
			NYSE17 & $0.25$  & $0.47$ & $0.22$ \\ 
			\hline 
			LSE & $0.13$ &  $0.69$ & $0.57$ \\ 
			\hline 
			FWB & $0.50$ &  $0.54$ & $0.30$ \\ 
			\hline 
			TSE & $0.38$ & $0.33$ & $0.33$ \\ 
			\hline 
			HKSE & $0.34$ &  $0.45$ & $0.32$ \\ 
			\hline 
			NYSE99 & $0.28$ & $0.60$ & $0.46$ \\ 
			\hline 
\end{tabular} 
\end{center}
	\caption{Linear-partial $K\tau$ correlation $K\tau_{par}= K\tau_{\hat B,  \bar \rho |\ln(mktcap)}$ between the residuals of average correlation $\bar\rho$ and $\hat B$ when they have been regressed against the logarithm of capitalization. All the p-values are less than 0.01 apart LSE, which is 0.03. The coefficients of determination $R^2_{\bar \rho}$ and $R^2_{\hat B}$ are also reported for the linear fit between the logarithm of the capitalization and respectively $\bar\rho$ and $\hat B$.}
	\label{tab_par_tau}
\end{table}

\subsection{Role of the Kurtosis}\label{subsec_kurtosis}
From an empirical perspective, as for the role of capitalization, it is possible to question if the kurtosis is playing a key role in this sort of analysis, leaving almost nothing to the multiscaling feature of the data. 
Kurtosis plays a key role in financial statistics. It is commonly used to analyse extreme events regarding the unconditional distribution of returns and to build risk assessment models, such as VaR and stress testing. Kurtosis of stock $i$'s log-returns is defined as:

\begin{equation}
K(r_i)=E\bigg[\bigg(\frac{r_i-\mu}{\sigma}\bigg)^4\bigg]=\frac{E[(r_i-\mu)^4]}{E[(r_i-\mu)^2]^2},
\end{equation}

 where $r_i$ is the $i$-th observation of the time series and $\mu$ and $\sigma$ are the mean and the standard deviation of the time series. Given its common use to detect extreme events, kurtosis is generally referred to as a measure of the fatness of the tails of a distribution, but having fatter tails is just one of the source of higher kurtosis. 
  This is because kurtosis depicts two different aspects of a distribution, namely the peakedness and taildeness. The first one refers to the fact that a distribution is more peaked on the centre while the second refers to the fatness of the tails. In order to have a higher kurtosis one needs to have a leptokurtic distribution, i.e. more peaked and with fatter tails \citep{kurtosis_decarlo, kurtosis_moors}. Although other researchers pointed out that the kurtosis coefficient is substantially more influenced by taildeness than by peakedness \citep{kurtosis_ruppert, kurtosis_westfall}, this double feature of kurtosis makes this measure not perfectly suitable to be used for tails and scaling analysis.

\begin{figure}[!h]
\begin{center}
		\includegraphics[width=0.70\textwidth]{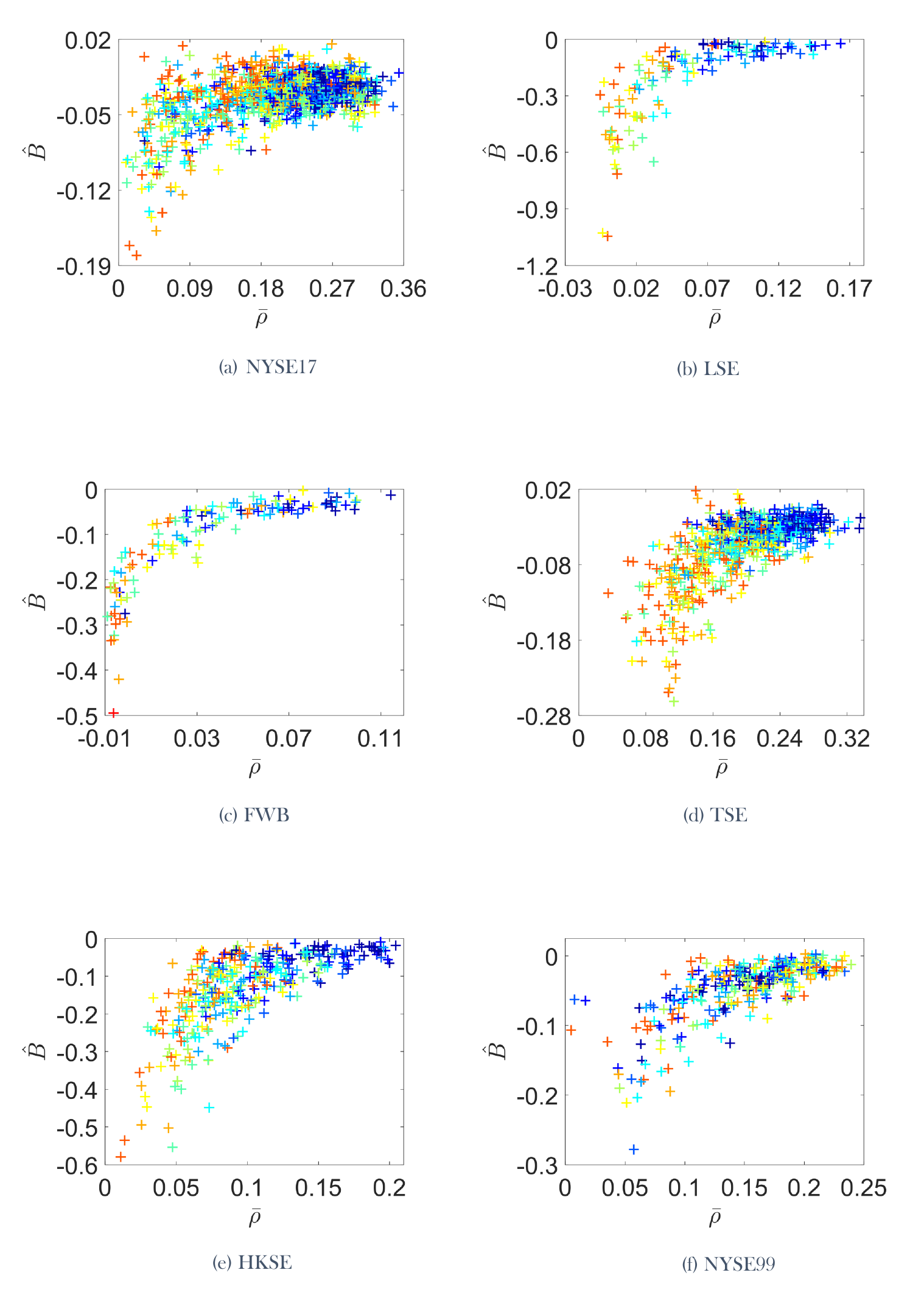}
\end{center}
	\caption{Empirical evidence of the dependence between the multiscaling properties, represented by the proxy $\hat B$ of a time-series and its average correlation $\bar \rho$. The colours from blue to red represent the increasing kurtosis.}
	\label{rho_VS_B_logK}
\end{figure}

In Fig.\ref{rho_VS_B_logK} we report the relationship between $\bar{\rho}$ and $\hat{B}$ where the colour of the points are, in this case, function of the kurtosis.\footnote{We use the logarithm of the kurtosis since the range of the measure across each market covers many orders of magnitudes.}
As it is possible to notice, a clear pattern is not depicted in the majority of markets. However, it is worth noting that in different cases, high level of kurtosis is matched with an high level of multiscaling.

As done in Subsec.\ref{subsec_cap}, we analyse the role of kurtosis in the observed stylized fact found in Subsec.\ref{subsec_main}. We investigate this relationship in three different ways. The first one is to check if the kurtosis exhibits the same clear stylized pattern with the average correlation as the multiscaling proxy does. The second one is to compare the multiscaling proxy with the kurtosis, to check for empirical overlapping information, stressing the fact that the two quantities reflect different aspects of the data, one being a feature of the unconditional distribution, while the other being related to the scaling exponent of the $q$-order moments when different time aggregations $\tau$ are taken into consideration. The third analysis we perform is, as we did with the capitalization in the previous subsection, to compute the linear-partial $K\tau$ between $\bar\rho$ and $\hat B$ after removing the linear relationship of the kurtosis from the two variables.\footnote{Filtering out the non-linear effect and both the linear and non-linear effect does not change the results.}

\begin{figure}[!h]
\begin{center}
		\includegraphics[width=0.70\textwidth]{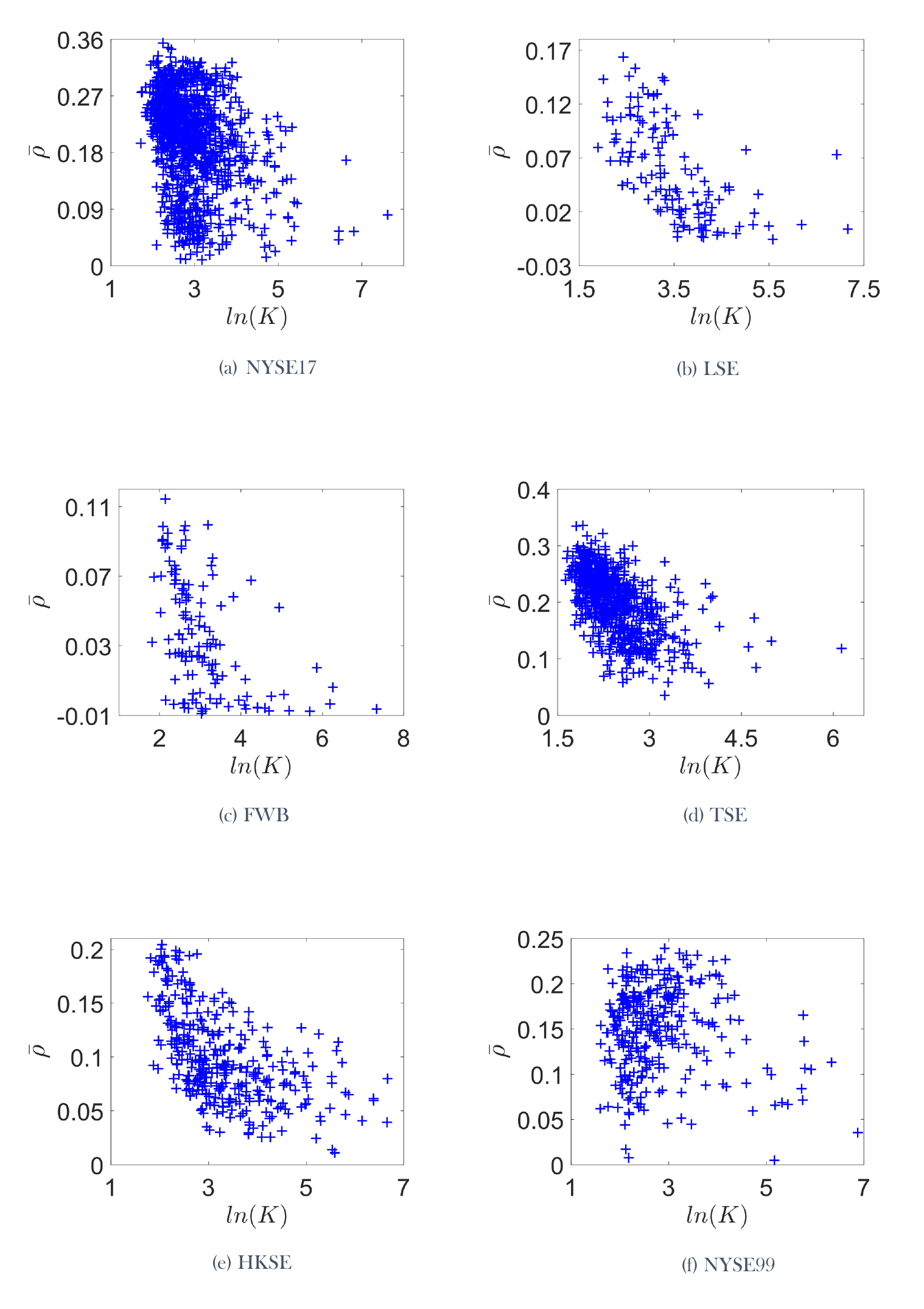}
\end{center}
	\caption{Empirical evidence of the dependence between the logarithm of kurtosis $\ln(K)$ of a log-returns time-series and its average correlation $\bar \rho$.}
	\label{K_VS_rhobar}
\end{figure}

Results for the first analysis are showed in Fig.\ref{K_VS_rhobar}. From the graphs some conclusions can be drawn. In particular, the clear pattern we got from the relationship between average correlation of the stocks and their multiscaling proxy is not present here. The relationship is slightly negative in some market and negligible in some others. In addition, it is possible to notice that the relationship between the two variables is very noisy.
This gives us two different interpretations. The first one is that for the markets in which the relationship is slightly negative, stocks with higher kurtosis turn out to have a lower average correlation with respect to the other stocks in the same market, while in the second case, no sensible conclusion can be made. This evidence suggests that the kurtosis is not a proper measure to put in relation with the average correlation for the markets under analysis.


\begin{figure}[!h]
\begin{center}
		\includegraphics[width=0.70\textwidth]{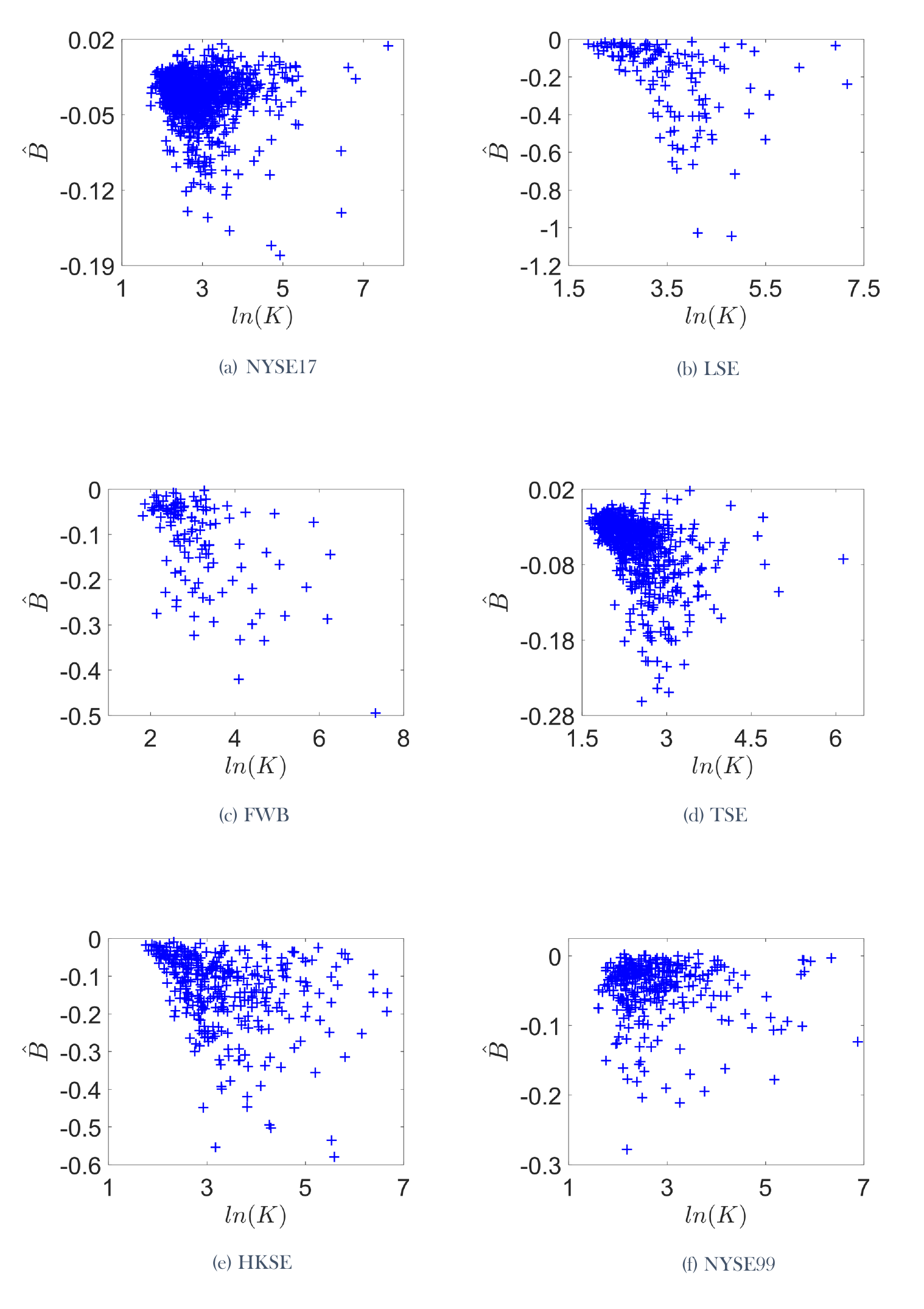}
\end{center}
	\caption{Empirical evidence of the dependence between the logarithm of kurtosis $\ln(K)$ of a time-series and its multiscaling proxy $\hat B$.}
	\label{K_VS_B}
\end{figure}

The second analysis regards the relationship between multiscaling proxy and the kurtosis coefficient of the log-returns' unconditional distribution. This analysis is carried out to understand if the two measures depict the same information of the data, leaving no point in measuring multiscaling. We report the scatter plots between the kurtosis coefficient and the multiscaling proxy in Fig.\ref{K_VS_B}. As it is possible to notice from the graphs, the two measures do not evidence a clear pattern among the six markets. This is in line with the fact that kurtosis is only a part of the multiscaling measurements, giving rise to the bias found in \cite{buonocore} when stocks with very high kurtosis are analysed.

The last investigation regards a quantitative assessment of what we presented in this subsection. In particular, we will rely on the linear-partial $K\tau$ correlation between $\bar \rho$ and $\hat B$ after the linear relationship between the kurtosis and the two variables is filtered out. Results of this analysis are reported in Tab.\ref{tab_par_tau_kurtosis}. As it can be noticed, the relationship remains strong and statistically significant at any level of significance for all the markets analysed.

\begin{table}[!h]
\begin{center}
\begin{tabular}{|c|c|c|c|}
			\hline 
			Market & $K\tau_{par}$ & $R^2_{\bar \rho}$ & $R^2_{\hat B}$ \\ 
			\hline 
			NYSE17 & $0.28$  & $0.11$ & $0.01$ \\ 
			\hline 
			LSE & $0.55$ &  $0.36$ & $0.17$ \\ 
			\hline 
			FWB & $0.56$ &  $0.25$ & $0.30$ \\ 
			\hline 
			TSE & $0.39$ & $0.31$ & $0.20$ \\ 
			\hline 
			HKSE & $0.49$ &  $0.32$ & $0.11$ \\ 
			\hline 
			NYSE99 & $0.53$ & $0.01$ & $0.01$ \\ 
			\hline 
\end{tabular} 
\end{center}
	\caption{Linear-partial $K\tau$ correlation $K\tau_{par}= K\tau_{\hat B,  \bar \rho |\ln(K)}$ between the residuals of average correlation $\bar\rho$ and $\hat B$ when they have been regressed against the logarithm of kurtosis. All the p-values are less than 0.01. The coefficients of determination $R^2_{\bar \rho}$ and $R^2_{\hat B}$ are also reported for the linear fit between the logarithm of the kurtosis and respectively $\bar\rho$ and $\hat B$.}
	\label{tab_par_tau_kurtosis}
\end{table}

\section{Discussion and conclusion}\label{sec_conclusion}
We find an empirical relationship which links a univariate property, i.e. the degree of multiscaling behaviour of a time series, to a multivariate one, i.e. the average correlation of the stock log-returns with the other stocks traded in the same market. Since the scaling measured for low aggregation horizon is biased by the presence of log-return tails and time autocorrelation \citep{buonocore}, we investigated which of the two gives the major contribution to the degree of multiscaling behaviour. In order to do so we used the shuffling technique to isolate the tails contribution and the normalization technique to isolate the autocorrelation contribution (see for example \citep{zhou,buonocore}). It turns out that the dependence found is due almost exclusively to the tails of log-returns distribution. However, we also found that either $\bar \rho$ and $\hat B$ are correlated with the logarithm of capitalization of the analysed stocks. In order to understand if the dependence from the logarithm of capitalization is fully explaining the stylized fact we discovered, we investigated the partial correlation between $\bar \rho$ and $\hat B$ using the logarithm of capitalization as a control variable. It turned out that removing the contribution due to the linear relation with the logarithm of capitalization of the two variables does not fully explain the non-linear dependence observed between $\bar \rho$ and $\hat B$. We also reproduced the same analysis by removing the non-linear dependence of the logarithm of capitalization from $\bar \rho$ and $\hat B$. We performed this by conditioning on the squared values of the logarithm of capitalization and by removing both the linear and non-linear effect of the logarithm of capitalization by a second order polynomial regression. In the first case, we find that the relationship remains statistically unchanged while in the second case, removing both linear and non-linear dependence, the result obtained for LSE market becomes statistically not significant, while the others, even if decreased in their strength, remain statistically significant at $1\%$ significance level.

We completed our set of robustness checks analysing the role of kurtosis in the same fashion we did with capitalization. We found that the role of kurtosis is marginal on the relationship between multiscaling and average correlation.
We interpret these findings as an evidence of the fact that the observed relationship must have a deeper origin. 

According to the observations in \citep{buonocore,buonocore2}, a shuffled empirical log-returns time series should scale asymptotically as a Brownian Motion due to the Central Limit Theorem \citep{buonocore, buonocore2}. The reason why the scaling at small aggregation regimes for discrete time series is not aligned to the asymptotic one has been found in the instability under aggregation of the empirical distribution of log-returns \citep{buonocore2}. We thus interpret the stocks with a value of $\hat B$ near to zero as more stable under aggregation, since the transient of the scaling differs less from the asymptotic one. Since the only stable distribution with finite variance is the Gaussian distribution, we find that the most capitalized and most correlated stocks are those which are less volatile. This observation gives a statistical motivation to the fact that highly capitalized stocks are those which are less risky from an investor point of view. In particular, if a stock is characterized by $\hat B$ tending to $0$ and $\hat A$ tending to $0.5$, it means that the stock's behaviour differs less from the behaviour of a Brownian motion than stocks with negative values of $\hat B$ and values of $\hat A$ larger than $0.5$. These results can be analysed in the spirit of portfolio construction, where asset managers build portfolios exploiting this stylized fact of the markets in terms of multiscaling and dependency structure. In particular, the portfolio builder would have to trade-off low correlation which means diversification with the multiscaling of the stocks, which renders each stock less stable at different investments horizon. On the other hand, low multicaling (or uniscaling) stocks are the ones more stable to aggregation, but at the cost of lower diversification. The results of this paper clarify the interplay between multiscaling and average correlation and calls for further investigation on the multiscaling of portfolios at different investment horizons.


\section*{Acknowledgements}
Many thanks to the two anonymous reviewers for their important comments that have greatly improved the manuscript. T.D.M. wishes to thank the ESRC Network Plus project 'Rebuilding macroeconomics'.

\appendix

\section{Data pre-processing procedure}\label{appendix_cleaning}
In this appendix we describe the data pre-processing procedure applied to our dataset. The main reason to apply a pre-processing procedure to the dataset is that we do not want to remove a stock just because it was not traded on few days during the investigated time interval. The pre-processing procedure fills gaps by dragging the last available price and by assuming that a missing record in the price time series corresponds to a zero log-return. At the same time we do not want to drag too many prices because a time series filled with many zeros would not be informative. In light of this we remove from our dataset the time series that are too short. The detailed procedure goes as follows:
\begin{enumerate}
\item Remove from the dataset the price time series with length less than $k$ times the longest one;
\item Find the common earliest day among the remaining time series;
\item Create a reference time series of dates when at least one of the stocks has been traded starting from the earliest common date found in the previous step;
\item Compare the reference time series of dates with the time series of dates of each stock and fill the gaps dragging the last available price.
\end{enumerate}
In this paper we chose $k=0.90$, however the results do not change if we pick a higher value of $k$ thus trying to keep as much as possible unprocessed time series.

\end{document}